\documentclass[onecolumn,nofootinbib]{revtex4}
\usepackage{graphicx}
\usepackage{latexsym}
\usepackage[colorlinks=true,linkcolor=red,citecolor=blue,urlcolor=green,dvipdfm]{hyperref}

\begin{document}

\title{Testing Dvali-Gabadadze-Porrati Gravity with Planck}

\author{Hong Li${}^{a,b}$}
\author{Jun-Qing Xia${}^{a}$}

\affiliation{${}^a$Key Laboratory of Particle Astrophysics, Institute of High Energy Physics, Chinese Academy of Science, P. O. Box 918-3, Beijing 100049, P. R. China}
\affiliation{${}^b$National Astronomical Observatories, Chinese Academy of Sciences, Beijing 100012, P. R. China}

\date{\today}

\begin{abstract}

Recently, the Planck collaboration has released the first cosmological papers providing the highest resolution, full sky, maps of the cosmic microwave background (CMB) temperature anisotropies. It is crucial to understand that whether the accelerating expansion of our universe at present is driven by an unknown energy component (Dark Energy) or a modification to general relativity (Modified Gravity). In this paper we study a phenomenological model which interpolates between the pure $\Lambda$CDM model and the Dvali-Gabadadze-Porrati (DGP) braneworld model with an additional parameter $\alpha$. Firstly, we calculate the ``distance information'' of Planck data which includes the ``shift parameter'' $R$, the ``acoustic scale'' $l_A$, and the photon decoupling epoch $z_\ast$ in different cosmological models and find that this information is almost independent on the input models we use. Then, we compare the constraints on the free parameter $\alpha$ of the DGP model from the ``distance information'' of Planck and WMAP data and find that the Planck data with high precision do not improve the constraint on $\alpha$, but give the higher median value and the better limit on the current matter density fraction $\Omega_m$. Then, combining the ``distance information'' of Planck measurement, baryon acoustic oscillations (BAO), type Ia supernovae (SNIa) and the prior on the current Hubble constant (HST), we obtain the tight constraint on the parameter $\alpha < 0.20$ at $95\%$ confidence level, which implies that the flat DGP model has been ruled out by the current cosmological data. Finally, we allow the additional parameter $\alpha < 0$ in our calculations and interestingly obtain $\alpha=-0.29\pm0.20$ ($68\%$ C.L.), which means the current data slightly favor the effective equation of state $w_{\rm eff}<-1$. More importantly, the tension between constraints on $H_0$ from different observational data has been eased.

\end{abstract}

\maketitle


\section{Introduction}\label{int}

Current cosmological observations, such as the cosmic microwave background (CMB) measurements of temperature anisotropies and polarization at high redshift $z\sim1090$ and the redshift-distance measurements of SNIa at $z<2$, have demonstrated that the universe is now undergoing an accelerated phase of expansion. The simplest explanation is that this behavior is driven by the cosmological constant or the dynamical dark energy models. On the other hand, this observed late-time acceleration of the expansion on the large scales could also caused by some modifications to general relativity.

One of well-known examples is the Dvali-Gabadadze-Porrati (DGP) braneworld model \cite{DGP}, in which the gravity leaks off the four dimensional Minkowski brane into the five dimensional bulk Minkowski space-time. In the framework of flat DGP model, the Friedmann equation will be modified as \cite{DGPCosmology}:
\begin{equation}
H^2-\frac{H}{r_c}=\frac{8\pi{G}}{3}\rho_m~,
\end{equation}
where $r_c=(H_0(1-\Omega_m))^{-1}$ is the crossover scale. This model realizes the so-called self-accelerating Universe that features a four-dimensional de Sitter phase. However, Recent studies have indicated that this self-accelerating branch of DGP model has been ruled out by the current observational data \cite{guo2006,maartens2006,yamamoto2006,rydbeck2007,amendola2008,fang2008,Fairbairn,Davis,Movahed,Song,
Sawicki,Zhu,Thomas,dgp_xia,DGPRev}. Furthermore, this self-accelerating solutions suffers the serious problem of ghost excitations \cite{nicolis04,koyama05,gorbunov06,charmousis06,deffayet06}. Even at the classical level, this theory is pathological.

In this paper we investigate an interesting phenomenological model, first introduced in Ref.\cite{mDGP}, which interpolates between the pure $\Lambda$CDM model and the DGP model with an additional parameter $\alpha$. Assuming the flatness of our universe, the Friedmann equation is modified as \cite{mDGP}:
\begin{equation}
H^2-\frac{H^\alpha}{r_c^{2-\alpha}}=\frac{8\pi{G}}{3}\rho_m~,
\end{equation}
where the crossover scale becomes $r_c=H^{-1}_0/(1-\Omega_m)^{\alpha-2}$. Thus, we can straightforwardly rewrite the above equation and obtain the expansion rate as following:
\begin{equation}
E^2(z)\equiv\frac{H^2}{H^2_0}=\Omega_m(1+z)^3+\frac{\delta{H^2}}{H^2_0}~,\label{ez}
\end{equation}
where the last term denotes the modification to the Friedmann equation of general relativity:
\begin{equation}
\frac{\delta{H^2}}{H^2_0}\equiv(1-\Omega_m)\frac{H^{\alpha}}{H^{\alpha}_0}=(1-\Omega_m)E^\alpha(z)~.\label{deltah}
\end{equation}
Here, $\alpha = 0$ and $\alpha=1$ denote the pure $\Lambda$CDM model and the DGP model, respectively.

Since the Planck collaboration has released the first cosmological papers providing the highest resolution, full sky, CMB maps \cite{planck_fit}, it is important to study the DGP model and revisit the constraint on parameters from the latest cosmological probes. In this paper we investigate this phenomenological DGP model and present the tight constraints from the latest Planck and WMAP9 data, the baryon acoustic oscillations (BAO) measurements from several large scale structure (LSS) surveys, the ``Union2.1'' compilation which includes 580 supernovae and the gaussian prior on the Hubble constant $H_0$. Our paper is organized as follows: In Section \ref{data} we describe the latest observational data sets used in the numerical analyses; Section \ref{result} contains our main global constraints on the phenomenological DGP model from the current observations. The last Section \ref{sum} is the conclusions.

\section{Observational Data}\label{data}

In our analysis, we consider the following cosmological probes: i) distance information of CMB measurements; ii) the baryon acoustic oscillation in the galaxy power spectra; iii) measurement of the current Hubble constant; iv) luminosity distances of type Ia supernovae.

\subsection{CMB Distance Information}

CMB measurement is sensitive to the distance to the decoupling epoch via the locations of peaks and troughs of the acoustic oscillations. Here we use the ``distance information", following the WMAP group \cite{Komatsu:2008hk}, which includes the ``shift parameter" $R$, the ``acoustic scale" $l_A$, and the photon decoupling epoch $z_\ast$. $R$ and $l_A$ correspond to the ratio of angular diameter distance to the decoupling era over the Hubble horizon and the sound horizon at decoupling, respectively, given by:
\begin{equation}
R=\frac{\sqrt{\Omega_mH^2_0}}{c}\chi(z_\ast)~,~~l_A=\frac{\pi\chi(z_\ast)}{\chi_s(z_\ast)}~,
\end{equation}
where $\chi(z_\ast)$ and $\chi_s(z_\ast)$ denote the comoving distance to $z_\ast$ and the comoving sound horizon at $z_\ast$, respectively. The decoupling epoch $z_\ast$ is given by Ref.\cite{zast}:
\begin{equation}
z_\ast=1048[1+0.00124(\Omega_b h^2)^{-0.738}][1+g_1(\Omega_m h^2)^{g_2}]~,
\end{equation}
where
\begin{equation}
g_1=\frac{0.0783(\Omega_b h^2)^{-0.238}}{1+39.5(\Omega_b
h^2)^{0.763}}~,~g_2=\frac{0.560}{1+21.1(\Omega_b h^2)^{1.81}}~.
\end{equation}
We calculate the likelihood of the CMB distance information as follows:
\begin{equation}
\chi^2=(x^{\rm th}_i-x^{\rm data}_i)(C^{-1})_{ij}(x^{\rm
th}_j-x^{\rm data}_j)~,
\end{equation}
where $x=(R,l_A,z_\ast)$ is the parameter vector and $(C^{-1})_{ij}$ is the inverse covariance matrix for the CMB distance information.

\begin{table}
\caption{Inverse covariance matrix for the distance information $l_A$, $R$ and $z_{\ast}$ from WMAP9 and Planck data in the pure $\Lambda$CDM framework.}\label{cmb_dis}
\begin{center}
\begin{tabular}{c|c|ccc}
\hline\hline
\multicolumn{5}{c}{WMAP9}\\
\hline
&Best fit&$l_A(z_{\ast})$&$R(z_{\ast})$&$z_{\ast}$\\
\hline
$l_A(z_{\ast})$ & $302.40$ & $3.182$ & $18.253$ & $-1.419$\\
$R(z_{\ast})$ & $1.7246$ &  & $11887.879$ & $-193.808$\\
$z_{\ast}$ & $1090.88$ &  & & $4.556$\\
\hline\hline
\multicolumn{5}{c}{Planck}\\
\hline
&Best fit&$l_A(z_{\ast})$&$R(z_{\ast})$&$z_{\ast}$\\
\hline
$l_A(z_{\ast})$ & $301.77$ & $44.077$ & $-383.927$ & $-1.941$\\
$R(z_{\ast})$ & $1.7477$ &  & $48976.330$ & $-630.791$\\
$z_{\ast}$ & $1090.25$ &  & & $12.592$\\
\hline\hline
\end{tabular}
\end{center}
\end{table}

\begin{figure}[t]
\begin{center}
\includegraphics[scale=0.35]{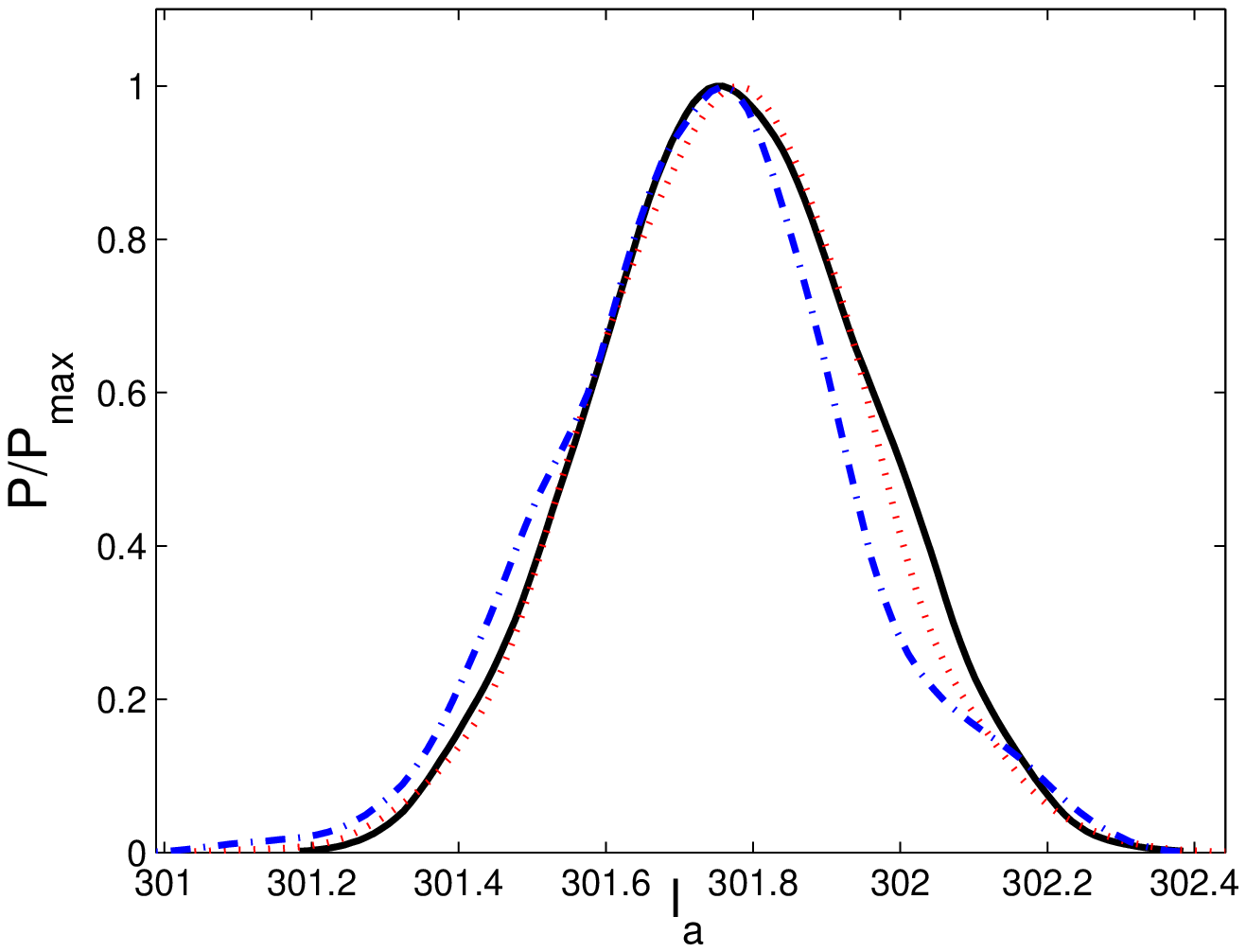}
\includegraphics[scale=0.35]{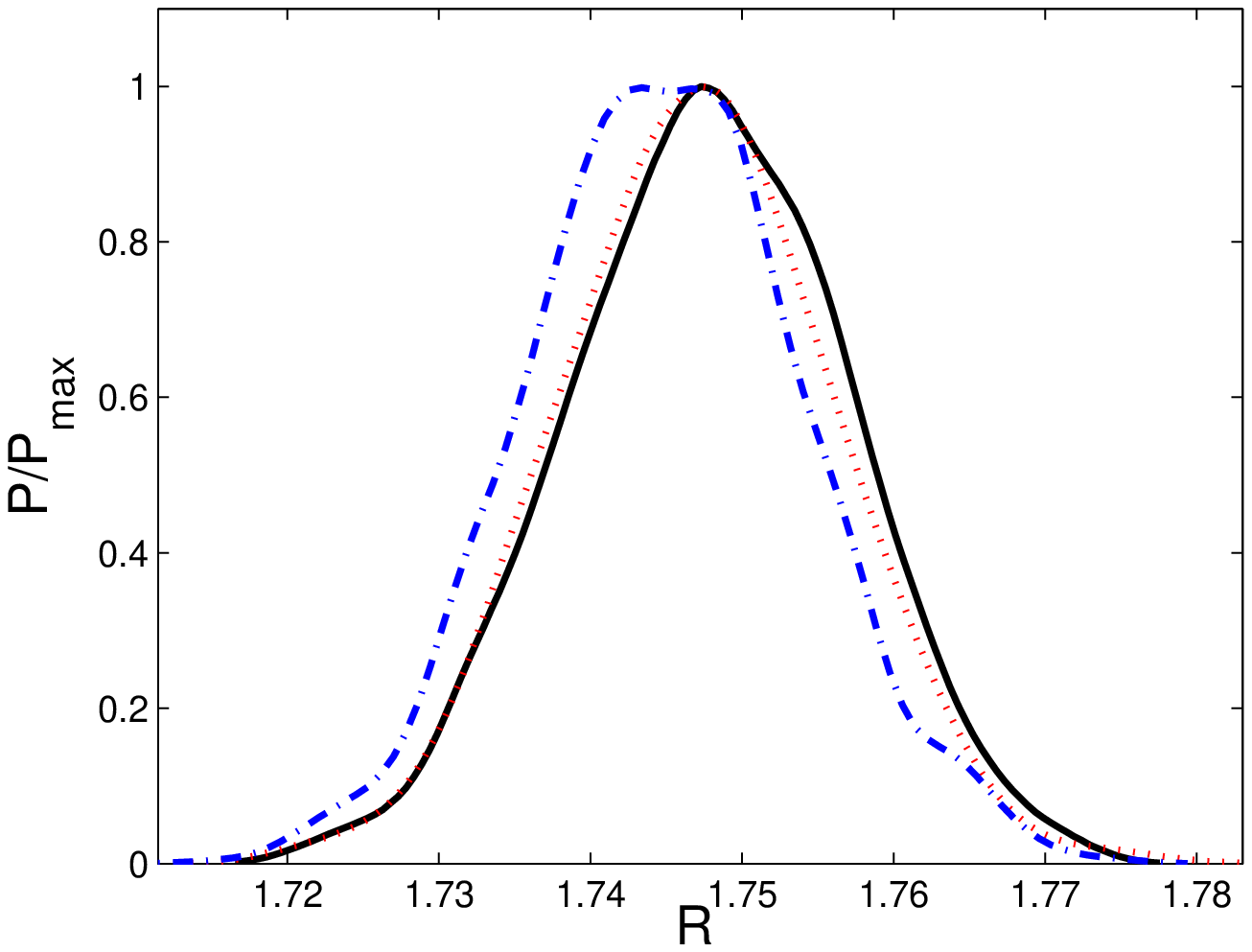}
\includegraphics[scale=0.35]{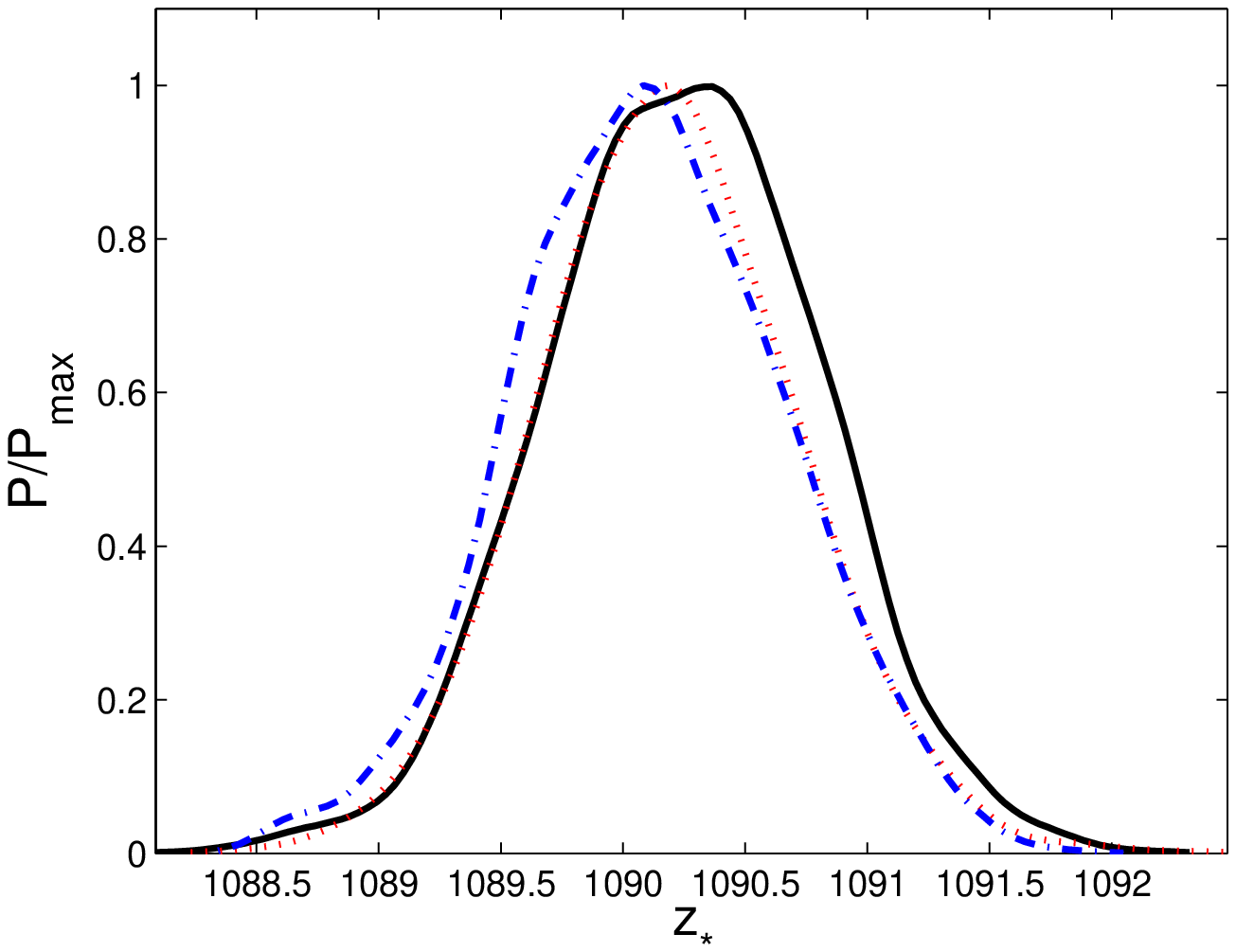}
\caption{One-dimensional posterior distributions of $l_A$, $R$, and $z_\ast$ obtained from the latest Planck data in different dark energy models: $\Lambda$CDM (black solid lines), the model with a constant equation of state (blue dash-dotted lines) and the model with a time-evolving equation of state (red dotted lines). \label{planck_dis}}
\end{center}
\end{figure}

In table \ref{cmb_dis} we show the inverse covariance matrix for the distance information from the WMAP9 and Planck measurements in the pure $\Lambda$CDM framework. Ref. \cite{Li_dis} has demonstrated that the distance information from the WMAP measurement is almost independent on the input dark energy models. Following the method described in Ref. \cite{Li_dis}, here we also check the distance information obtained from the Planck measurement in different input dark energy models, shown in figure \ref{planck_dis}. We find that, the distributions of the distance priors given by the Planck temperature power spectrum are almost the same in three different dark energy models: the standard $\Lambda$CDM, the dark energy models with a constant equation of state ($w$CDM) or a time-evolving equation of state ($w(a)=w_0+w_a(1-a)$ \cite{cpl}) ($w(z)$CDM). In table \ref{dp_num} we list constraints on $l_A$, $R$ and $z_{\ast}$ from the Planck data in different input dark energy models. Therefore, in our calculations we use the distance information of WMAP9 and Planck measurements obtained in the pure $\Lambda$CDM model to constrain the DGP model.

\begin{table}
\caption{Constraints on the distance information $l_A$, $R$ and $z_{\ast}$ from Planck data in different dark energy models.}\label{dp_num}
\begin{center}
\begin{tabular}{c|ccc}
\hline\hline

&$l_A(z_{\ast})$&$R(z_{\ast})$&$z_{\ast}$\\
\hline

$\Lambda$CDM & $301.77\pm0.18$ & $1.7477\pm0.0091$ & $1090.25\pm0.55$ \\
$w$CDM & $301.76\pm0.18$ &  $1.7470\pm0.0090$ & $1090.19\pm0.52$\\
$w(z)$CDM &$301.74\pm0.19$ &  $1.7448\pm0.0090$ & $1090.11\pm0.54$\\
\hline\hline
\end{tabular}
\end{center}
\end{table}

\subsection{Other Measurements}

Baryon Acoustic Oscillations provides an efficient method for measuring the expansion history by using features in the clustering of galaxies within large scale surveys as a ruler with which to measure the distance-redshift relation. It provides a particularly robust quantity to measure \cite{bao}. It measures not only the angular diameter distance, $D_A(z)$, but also the expansion rate of the universe, $H(z)$, which is powerful for studying dark energy \cite{task}. Since the current BAO data are not accurate enough for extracting the information of $D_A(z)$ and $H(z)$ separately \cite{okumura}, one can only determine an effective distance \cite{baosdss}:
\begin{equation}
D_V(z)=[(1+z)^2D_A^2(z)cz/H(z)]^{1/3}~.
\end{equation}
Following the Planck analysis \cite{planck_fit}, in this paper we use  the BAO measurement from the 6dF Galaxy Redshift Survey (6dFGRS) at a low redshift ($r_s/D_V (z = 0.106) = 0.336\pm0.015$) \cite{6dfgrs}, and the measurement of the BAO scale based on a re-analysis of the Luminous Red Galaxies (LRG) sample from Sloan Digital Sky Survey (SDSS) Data Release 7 at the median redshift ($r_s/D_V (z = 0.35) = 0.1126\pm0.0022$) \cite{sdssdr7}, and the BAO signal from BOSS CMASS DR9 data at ($r_s/D_V (z = 0.57) = 0.0732\pm0.0012$) \cite{sdssdr9}.

In our analysis, we add a Gaussian prior on the current Hubble constant given by Ref.\cite{hst_riess}; $H_0 = 73.8 \pm 2.4$ ${\rm km\,s^{-1}\,Mpc^{-1}}$ ($68\%$ C.L.). The quoted error includes both statistical and systematic errors. This measurement of $H_0$ is obtained from the magnitude-redshift relation of 240 low-z Type Ia supernovae at $z < 0.1$ by the Near Infrared Camera and Multi-Object Spectrometer (NICMOS) Camera 2 of the Hubble Space Telescope (HST). This is a significant improvement over the previous prior, $H_0 = 72 \pm 8$ ${\rm km\,s^{-1}\,Mpc^{-1}}$, which is from the Hubble Key project final result. In addition, we impose a weak top-hat prior on the Hubble parameter: $H_0 \in [40, 100]$ ${\rm km\,s^{-1}\,Mpc^{-1}}$.

Finally, we include data from Type Ia supernovae, which consists of luminosity distance measurements as a function of redshift. In this paper we use the latest SN data sets from the Supernova Cosmology Project, ``Union Compilation 2.1'', which consists of 580 samples and spans the redshift range $0\lesssim{z}\lesssim1.55$ \cite{union}. This data set also provides the covariance matrix of data with and without systematic errors. In order to be conservative, we use the covariance matrix with systematic errors. When calculating the likelihood from SN, we marginalize over the absolute magnitude M, which is a nuisance parameter, as done in Ref.\cite{SNMethod}.

\section{Numerical Results}\label{result}

In our analysis, we perform a global fitting using the {\tt CosmoMC} package \cite{cosmomc}, a Monte Carlo Markov chain (MCMC) code, which has been modified to calculate the background evolution of this phenomenological DGP model. We vary the following cosmological parameters with top-hat priors: the cold dark matter energy density parameter $\Omega_c h^2 \in [0.01, 0.99]$, the baryon energy density parameter $\Omega_b h^2 \in [0.005, 0.1]$, the current Hubble constant $H_0 \in [40,100]$ ${\rm km\,s^{-1}\,Mpc^{-1}}$ and the additional parameter $\alpha$ in this phenomenological DGP model.



\begin{figure}[t]
\begin{center}
\includegraphics[scale=0.45]{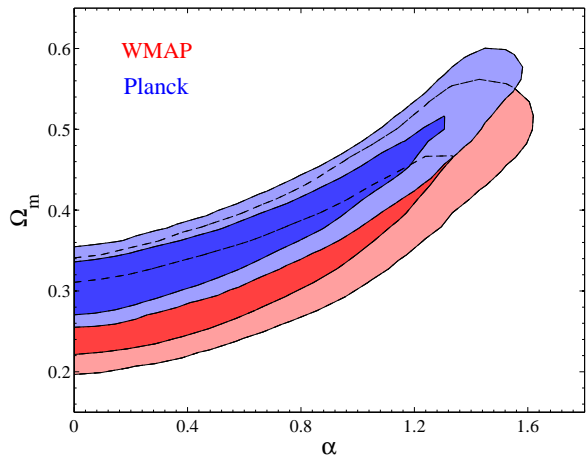}
\caption{Two-dimensional contours in the ($\alpha,\Omega_m$) panel obtained from the WMAP9 (red) and Planck (blue) data.\label{cmb_alone}}
\end{center}
\end{figure}

Firstly, we consider the constraint on the DGP models from the Planck and WMAP9 data alone. In Figure \ref{cmb_alone} we show the two-dimensional contours in the ($\alpha,\Omega_m$) panel. As we know, the CMB anisotropies mainly contain the information about the high-redshift universe, but it is not directly sensitive to lower-redshift phenomena, such as the nature of accelerating Universe. Therefore, CMB data alone can not constrain the parameter $\alpha$ of the DGP model very well. WMAP9 and the more accurate Planck data almost give the identical constraint on $\alpha$, namely the $95\%$ C.L. upper limit is $\alpha < 1.39$. The DGP model ($\alpha=1$) and the $\Lambda$CDM model ($\alpha=0$) can not be distinguished by the CMB data alone. We need to add some extra information from the low-redshift probes to break the degeneracy. Interestingly, the two-dimensional contours in Figure \ref{cmb_alone} do not overlap totally. The Planck data prefer a higher value of $\Omega_m$ that that obtained from the WMAP9 data, due to the higher value of its ``shift parameter" $R\sim\sqrt{\Omega_m}$. This tension was also found by the Planck group \cite{planck_fit} and has been widely discussed in the literature \cite{verde,waynehu,xia_tension}.

\begin{figure}[t]
\begin{center}
\includegraphics[scale=0.45]{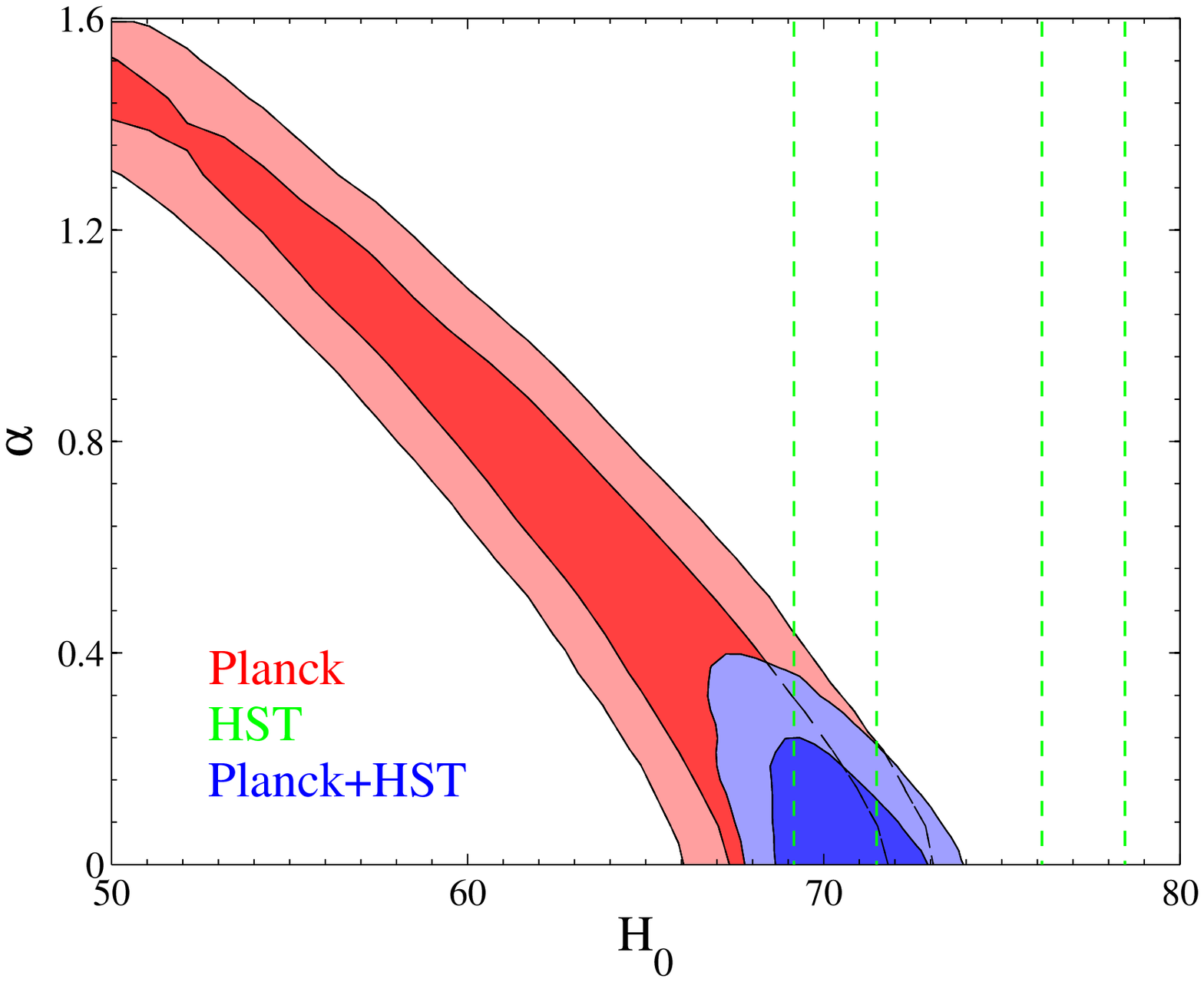}
\includegraphics[scale=0.45]{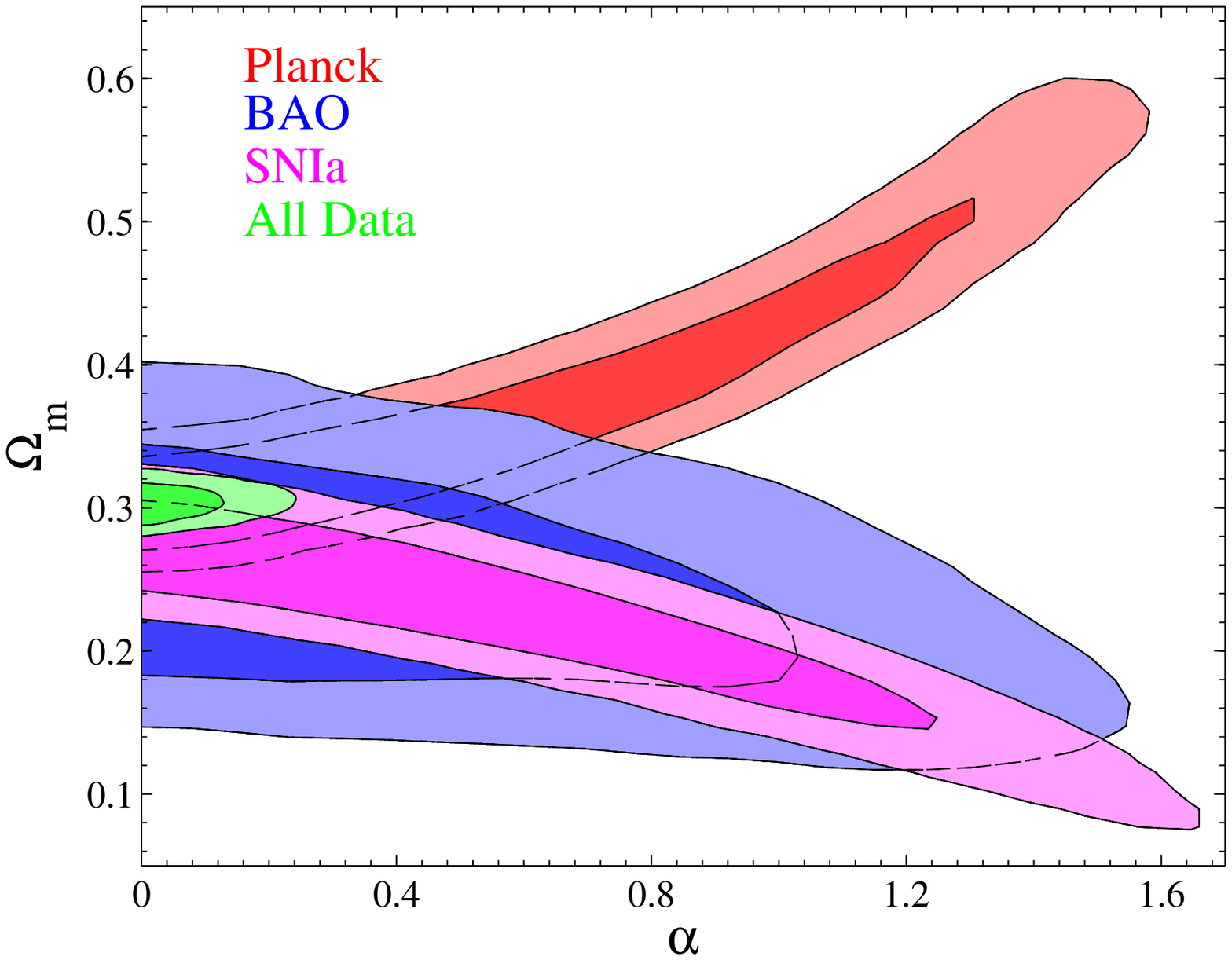}
\caption{Left Panel: Two-dimensional contours in the ($H_0,\alpha$) panel obtained from the Planck (red) and Planck+HST (blue). Four green vertical dashed lines denote the $1,2\,\sigma$ limits of the HST gaussian prior we use. Right Panel: Two-dimensional contours in the ($\alpha,\Omega_m$) panel obtained from different data combinations: Planck (red), BAO (blue), SNIa (magenta) and all data together (green). \label{planck}}
\end{center}
\end{figure}

This tension is also shown in the constraint of $H_0$. The Planck data alone can only yield a very weak constraint on the Hubble constant: $H_0 < 70.0$ ${\rm km\,s^{-1}\,Mpc^{-1}}$ ($95\%$ C.L.), see the left panel of Figure \ref{planck}. This result is apparently lower than the HST gaussian prior: $H_0 = 73.8 \pm 2.4$ ${\rm km\,s^{-1}\,Mpc^{-1}}$ (68\% C.L.), which is consistent with that found by the Planck group. Due to the strong degeneracy between $\alpha$ and $H_0$, the flat DGP model ($\alpha=1$) requires a low value of $H_0$, in order to produce the same value of $R$. Therefore, adding HST prior significantly increases the $\chi^2$ of this model. When we forcibly add the HST prior into the calculation, the joint constraint prefers a higher value of $H_0$, $H_0 = 70.0 \pm 1.4$ ${\rm km\,s^{-1}\,Mpc^{-1}}$ (68\% C.L.), and the parameter $\alpha$ is tightly limited, $\alpha<0.32$ ($95\%$ C.L.).

Besides the direct $H_0$ probe, we also consider some other low-redshift probes, like BAO and SNIa, in our analysis. In the right panel of Figure \ref{planck} we show the constraints on $\alpha$ from BAO (blue) and SNIa (magenta), respectively, which are similar with that obtained from the Planck data. However, different from the degeneracy between $\Omega_m$ and $\alpha$ in the Planck data (red), $\Omega_m$ and $\alpha$ are strongly anti-correlated. The reason for this degeneracy is that the constraint mainly comes from the luminosity and angular diameter distance information. From eqs.(\ref{ez}) and (\ref{deltah}) we can see that when $\alpha$ is increased, the contribution of the last $\alpha$ term to the expansion rate will become large, due to the positive $E(z)$. Consequently, $\Omega_m$ must be decreased correspondingly in order to produce the same expansion rate. Therefore, when we combine the BAO or SNIa and the Planck data, the joint constraints are significantly shrunk, namely the $95\%$ upper limits are $\alpha<0.60$ and $\alpha<0.23$ from Planck+BAO and Planck+SNIa, respectively. When we combine all these data together (green), the constraint on $\alpha$ becomes tighter further,
\begin{equation}
\alpha<0.20~~(95\%\,{\rm C.L.})~.
\end{equation}
The flat DGP model ($\alpha=1$) has been ruled out with very high significance, which is consistent with other works (see e.g. refs. \cite{dgp_xia,DGPRev}).

\begin{figure}[t]
\begin{center}
\includegraphics[scale=0.45]{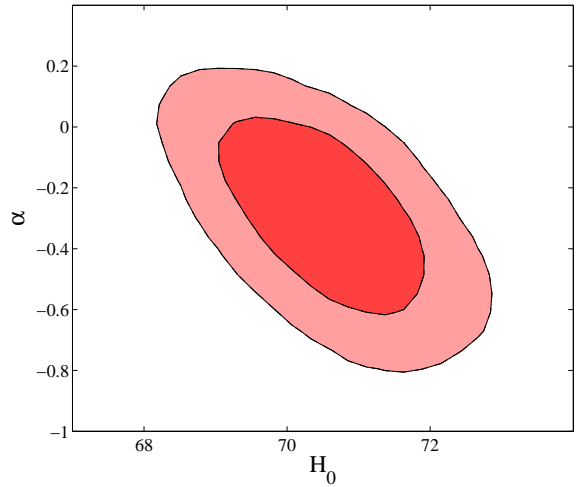}
\caption{Two-dimensional contours in the ($H_0,\alpha$) panel obtained from all data sets together (red). \label{alpha_less}}
\end{center}
\end{figure}

Until now, we only consider the phenomenological DGP model with $\alpha \ge 0$. Based on eq.(\ref{ez}), we can obtain the effective equation of state (EoS):
\begin{equation}
w_{\rm eff}(z)=-1+\frac{\alpha}{3}(1+z)\frac{E'(z)}{E(z)}~,
\end{equation}
where the prime denotes the derivative with respect to the redshift $z$. Therefore, $\alpha \ge 0$ corresponds to the effective EoS $w_{\rm eff}(z) \ge -1$ at high redshifts. Interestingly, there is a possibility where the parameter $\alpha$ is less than zero. In this case, the effective EoS can be more negative than $w\equiv-1$, without violating the weak-energy condition \cite{mDGP}. Thus, we extend our previous analyses and allow the parameter $\alpha < 0$. In figure \ref{alpha_less} we show the two-dimensional contour in the ($H_0,\alpha$) pane obtained from all data sets together. The data yield the tight constraint on the additional parameter of
\begin{equation}
\alpha = -0.29 \pm 0.20~~(68\%~{\rm C.L.})~,
\end{equation}
which implies that the pure $\Lambda$CDM model ($\alpha=0$) is consistent with the data, but the model with a negative value of $\alpha$ is slightly favored. Based on the equation of $w_{\rm eff}$, we can see that the current observational data favor the effective EoS of dark energy $w_{\rm eff} < -1$, which is consistent with previous works \cite{xia_tension}. Since $\alpha$ and $H_0$ are anti-correlated, consequently, a high value of the Hubble constant is also obtained: $H_0=71.0\pm0.9$ ${\rm km\,s^{-1}\,Mpc^{-1}}$ ($68\%$ C.L.), which is  consistent with the direct probe of $H_0$ from the HST measurement \cite{hst_riess}.

\section{Summary}\label{sum}

As an alternative approach to generate the late-time acceleration of the expansion of our Universe, models of modifications of gravity have attracted a lot of interests in the phenomenological studies recently. In this paper we investigate an interesting phenomenological model which interpolates between the pure $\Lambda$CDM model and the flat DGP braneworld model with an additional parameter $\alpha$.

We find that the CMB data alone can not give tight constraint on $\alpha$, due to the strong degeneracies among $\Omega_m$, $H_0$ and $\alpha$. WMAP9 and the more accurate Planck data almost give the identical constraint, $\alpha < 1.39$ ($95\%$ C.L.). But Planck data give a higher value of $\Omega_m$ than that from WMAP9 data, which is similar with the tension found by the Planck group. When we add the HST prior, BAO or SNIa, the constraint on $\alpha$ becomes significantly stringent. Combining all data together, we obtain the tightest constraint on the parameter $\alpha < 0.20$ at $95\%$ confidence level, which implies that the flat DGP model ($\alpha=1$) is incompatible with the current observations, while the pure $\Lambda$CDM model still fits the data very well. We also allow the additional parameter $\alpha < 0$ in our calculations, which corresponds to the effective equation of state $w_{\rm eff}<-1$. Our result shows that the current data slightly favor a negative value: $\alpha=-0.29\pm0.20$ ($68\%$ C.L.). Consequently, the obtained constraint on $H_0$ is consistent with the direct probe of $H_0$ from the HST. The tension on $H_0$ from different observational data found by the Planck collaboration \cite{planck_fit} has been eased.

\section*{Acknowledgements}

HL is supported in part by the National Science Foundation of China under Grant Nos. 11033005, by the 973 program under Grant No. 2010CB83300, by the Chinese Academy of Science under Grant No. KJCX2-EW-W01. JX is supported by the National Youth Thousand Talents Program and the grants No. Y25155E0U1 and No. Y3291740S3.



\begin{thebibliography}{nn}

\bibitem{DGP}
G.~R.~Dvali, G.~Gabadadze and M.~Porrati, Phys.\ Lett.\ B {\bf 485}, 208 (2000).

\bibitem{DGPCosmology}
C.~Deffayet, Phys.\ Lett.\ B {\bf 502}, 199 (2001).

\bibitem{mDGP}
G.~Dvali and M.~S.~Turner, arXiv:astro-ph/0301510.

\bibitem{Song}
Y. S. Song, Phys. Rev. D 71, 024026 (2005).
\bibitem{Fairbairn}
M. Fairbairn and A. Goobar, Phys. Lett. B 642, 432 (2006).
\bibitem{guo2006}
Z. K. Guo, Z. H. Zhu, J. S. Alcaniz and Y. Z. Zhang, Astrophys. J. 646, 1 (2006).
\bibitem{maartens2006}
R. Maartens and E. Majerotto, Phys. Rev. D 74, 023004 (2006).
\bibitem{yamamoto2006}
K. Yamamoto, B. A. Bassett, R. C. Nichol and Y. Suto, Phys. Rev. D 74, 063525 (2006).
\bibitem{Davis}
T. M. Davis et al., Astrophys. J. 666, 716 (2007).
\bibitem{rydbeck2007}
S. Rydbeck, M. Fairbairn and A. Goobar, JCAP 0705, 003 (2007).
\bibitem{Sawicki}
Y. S. Song, I. Sawicki, and W. Hu, Phys. Rev. D 75, 064003 (2007).
\bibitem{Zhu}
Z. H. Zhu and M. Sereno, Astron. Astrophys. 487, 831 (2008).
\bibitem{amendola2008}
L. Amendola, M. Kunz and D. Sapone, JCAP 0804, 013 (2008).
\bibitem{fang2008}
W. Fang, S. Wang, W. Hu, Z. Haiman, L. Hui and M. May, Phys. Rev. D 78, 103509 (2008).
\bibitem{Movahed}
M. S. Movahed, M. Farhang, and S. Rahvar, Int. J. Theor. Phys. 48, 1203 (2009).
\bibitem{Thomas}
S. A. Thomas, F. B. Abdalla, and J. Weller, Mon. Not. Roy. Astron. Soc. 395, 197 (2009).

\bibitem{dgp_xia}
J.-Q.~Xia, Phys. Rev. {\bf D79}, 103527 (2009).

\bibitem{DGPRev}
A.~Lue, Phys.\ Rept.\  {\bf 423}, 1 (2006); R.~Durrer and R.~Maartens, arXiv:0811.4132, and also references therein.

\bibitem{nicolis04}
A.~Nicolis and R.~Rattazzi, JHEP {\bf 0406}, 059 (2004).
\bibitem{koyama05}
K. Koyama, Phys. Rev. D 72, 123511 (2005).
\bibitem{gorbunov06}
D. Gorbunov, K. Koyama and S. Sibiryakov, Phys. Rev. D 73, 044016 (2006).
\bibitem{charmousis06}
C. Charmousis, R. Gregory, N. Kaloper and A. Padilla, JHEP 0610, 066 (2006).
\bibitem{deffayet06}
C. Deffayet, G. Gabadadze and A. Iglesias, JCAP 0608, 012 (2006).

\bibitem{planck_fit}
Planck Collaboration, P. A. R. Ade, {\it et al.}, ArXiv e-prints (2013), arXiv:1303.5076.

\bibitem{Komatsu:2008hk}
E.~Komatsu, {\it et al.}, Astrophys.\ J.\ Suppl.\ Ser.\ {\bf 180}, 330 (2009).

\bibitem{zast}
W.~Hu and N.~Sugiyama, Astrophys.\ J.\ {\bf 471}, 542 (1996).

\bibitem{Li_dis}
H.~Li, J.-Q.~Xia, G.-B.~Zhao, Z.~Fan and X.~Zhang, Astrophys.\ J.\ {\bf 683}, L1 (2008).

\bibitem{cpl}
M. Chevallier and D. Polarski, Int. J. Mod. Phys. {\bf D10}, 213 (2001); E. V. Linder, Phys. Rev. Lett. {\bf 90}, 091301 (2003).

\bibitem{bao}
D. J. Eisenstein, H.-J. Seo and M. J. White, Astrophys. J. {\bf 664}, 660 (2007).

\bibitem{task}
A. Albrecht, {\it et al.}, ArXiv e-prints (2006), arXiv:astro-ph/0609591.

\bibitem{okumura}
T. Okumura, {\it et al.}, Astrophys. J. {\bf 676}, 889 (2008).

\bibitem{baosdss}
D. J. Eisenstein, {\it et al.}, Astrophys. J. {\bf 633}, 560 (2005).

\bibitem{6dfgrs}
F. Beutler, {\it et al.}, Mon. Not. Roy. Astron. Soc. {\bf 416}, 3017 (2011).

\bibitem{sdssdr7}
N. Padmanabhan, {\it et al.}, Mon. Not. Roy. Astron. Soc. {\bf 427}, 2132 (2012).

\bibitem{sdssdr9}
L. Anderson, {\it et al.}, Mon. Not. Roy. Astron. Soc. {\bf 428}, 1036 (2013).

\bibitem{hst_riess}
A. G. Riess, {\it et al.}, Astrophys. J. {\bf 730}, 119 (2011).

\bibitem{union}
N.~Suzuki {\it et al.}, Astrophys. J. {\bf 746}, 85 (2012).

\bibitem{SNMethod}
E.~Di~Pietro and J.~F.~Claeskens, Mon.\ Not.\ Roy.\ Astron.\ Soc.\ {\bf 341}, 1299 (2003).

\bibitem{cosmomc}
A. Lewis and S. Bridle, Phys. Rev. {\bf D66}, 103511 (2002); http://cosmologist.info/cosmomc/.

\bibitem{verde}
L.~Verde, P.~Protopapas and R.~Jimenez, ArXiv e-prints (2013), arXiv:1306.6766.

\bibitem{waynehu}
M.~Wyman, D.~H.~Rudd, R.~A.~Vanderveld and W.~Hu, ArXiv e-prints (2013), arXiv:1307.7715.

\bibitem{xia_tension}
J.-Q.~Xia, H.~Li and X.~Zhang, ArXiv e-prints (2013), arXiv:1308.0188.

\end{thebibliography}
\end{document}